\documentclass[prc,aps,floatfix,groupedaddress,amsmath,amssymb,twocolumn]{revtex4}

\usepackage{graphicx}
\usepackage{dcolumn}
\usepackage{bm}
\usepackage{color}

\usepackage[section]{placeins}
\usepackage{graphicx,epsfig}
\usepackage{amsmath}
\usepackage{amsfonts}
\usepackage{braket}
\usepackage{fancyhdr}
\usepackage{hyperref}
\usepackage[utf8]{inputenc}
\usepackage[T1]{fontenc}
\usepackage{amsthm}
\usepackage{hhline}
\usepackage{multirow}
\usepackage[figuresleft]{rotating}
\usepackage{gensymb}
\def\beq{\begin{equation}}
\def\eeq{\end{equation}}
\def\beqa{\begin{eqnarray}}
\def\eeqa{\end{eqnarray}}

\begin{document}

\title{Structure of single $\Lambda$-hypernuclei with Gogny-type $\Lambda$-nucleon forces}

\author{C.V. Nithish Kumar$^{1,2}$, L. M. Robledo$^{2,3}$ and I. Vida\~na$^4$}
\affiliation{$^1$Dipartimento di Fisica e Astronomia ``Ettore Majorana'', Universit\`a di Catania, Via Santa Sofia 64, I-95123 Catania, Italy }
\affiliation{$^2$Departamento de F\'{i}sica Te\'orica and CIAFF, Universidad Autónoma de Madrid, E-28049 Madrid, Spain}
\affiliation{$^3$Center for Computational Simulation, Universidad Polit\'ecnica de Madrid, Campus de Monteganceldo, Bohadilla del Monte, E-28660 Madrid, Spain}
\affiliation{$^4$Istituto Nazionale di Fisica Nucleare, Sezione di Catania, Via Santa Sofia 64, I-95123 Catania, Italy }


\begin{abstract}
We study the structure of single $\Lambda$-hypernuclei using the Hartree--Fock--Bogoliubov method. Finite range Gogny-type forces are used to describe the nucleon-nucleon and $\Lambda$-nucleon interactions. Three different $\Lambda$-nucleon Gogny
forces are built. The unknown parameters of these forces are obtained by fitting the experimental binding energies of the $1s$ 
$\Lambda$ single-particle state in various hypernuclei using the ``Simulated Annealing Method''. These forces are then used to calculate the binding energies of the other ($1p, 1d, 1f, 1g$) $\Lambda$ single-particle states in the different hypernuclei. The predicted values are found to be in good agreement with the experimental data for the three forces constructed. In addition, we calculate also the root-mean-square radii of ground state $\Lambda$ orbital, as well as
several global properties of the hypernuclei considered such as their ground-state Hartree--Fock--Bogoliubov energy, their pairing energy and their quadrupole moment.
\end{abstract}
\maketitle


\section{Introduction}
\label{sec:intro}

One of the aims of hypernuclear physics \cite{Botta12,Gal16,Hiyama18} is to relate the hyperon-nucleon (YN) and hyperon-hyperon (YY) interactions, poorly constrained by the limited amount of scattering data, with hypernuclear observables. Therefore, understanding the production, structure and decay mechanisms of hypernuclei is crucial to obtain information on these interactions and, in fact, it has been subject of intense theoretical and experimental research since  the first hyperfragment was observed by Danysz and Pniewski in a balloon-flown emulsion stack \cite{Danysz53} in 1952. Pion and proton beam production reactions in emulsions and later in $^4$He bubble chambers followed these initial cosmic-ray observations of hypernuclei, and allowed the identification and the determination of the binding energies, spins and lifetimes of single $\Lambda$-hypernuclei with $A\leq 15$ \cite{Juric73,Davis91} from the analysis of their mesonic weak decay. Average properties of heavier systems were estimated from spallation experiments, and two double-$\Lambda$ hypernuclei were reported from $\Xi^-$ capture \cite{Danysz63,Danysz63b,Prowse66,Dalitz89,Aoki91,Dover91,Franklin95}. Systematic investigations of hypernuclei started with the use of high-energy accelerators and modern electronic counters. In particular, the use of kaon, pion and electron beams has allowed the production of single $\Lambda$-hypernuclei through $(K^-,\pi^-)$ strangeness exchange reactions,
($\pi^+,K^+$) associate production reactions, and ($e,e'K^+$) electro-production reactions. A promising new way to produce hypernuclei by using stable and unstable heavy-ion beams was proposed a few years ago by the HypHI Collaboration at FAIR/GSI \cite{Bianchin09} which in a first experiment using a $^6$Li beam on a $^{12}$C target at 2AGeV led to the observation of the $\Lambda$ hyperon, and
the $^3_\Lambda$H and $^4_\Lambda$H hypernuclei \cite{Rappold13}. Nowadays, more than 40 single $\Lambda$-hypernuclei \cite{Hashimoto06} and a few double $\Lambda$  \cite{dl0,dl1,dl1b,dl2,dl3,dl4,dl5,nagara} and single $\Xi$ \cite{khaustov00,nakazawa15} ones have been identified. The existence of single $\Sigma$-hypernuclei has not been experimentally confirmed yet without ambiguity \cite{bertini80,bertini84,bertini85,piekarz82,yamazaki85,tang88,bart99,hayano89,nagae98}, suggesting that the $\Sigma$-nucleon interaction is most probably repulsive  \cite{dgm89,batty1,batty2,batty3,mares95,dabrowski99,noumi02,saha04,harada05,harada06}. 

Many of the theoretical approaches used to study the properties of single $\Lambda$-hypernuclei are based on a simple picture in which the hypernucleus is viewed as an ordinary nucleus plus a $\Lambda$ hyperon sitting in a single-particle state of an effective $\Lambda$-nucleus mean field potential. Among these approaches Woods--Saxon $\Lambda$-nucleus effective potentials, for instance, have been traditionally employed to describe reasonably well the measured hypernuclear single-particle states from medium to heavy hypernuclei \cite{ws1,ws1b,ws2,ws3,skhf1}. Non-relativistic Hartree--Fock calculations of hypernuclei with Skyrme-type YN interactions including density dependent effects and non-localities have been extensively performed improving the overall description of the
$\Lambda$ single-particle binding energies \cite{skhf2,skhf3,skhf4,skhf4b,skhf5,skhf5b,skhf5c,skhf5d,skhf6,skhf7,skhf8,skhf9}. Properties of hypernuclei have been also studied within a relativistic framework, such as Dirac phenomenology, where the $\Lambda$-nucleus potential is derived from the nucleon--nucleus one \cite{dirac1,dirac2}, or relativistic mean field (RMF) 
theory \cite{rmf1,rmf2,rmf3,rmf4,rmf5,rmf6,rmf7,rmf8,rmf8b,rmf9}. Microscopic calculations of the hypernuclear structure existing in the literature \cite{Yamamoto85,Yamamoto90,Yamamoto92,Yamamoto94,Halderson93,Hao93,Hjorth-Jensen96,Vidana97,Haidenbauer20} have been based on the construction of an effective YN interaction (G-matrix) obtained from the bare YN one by solving the 
Bethe--Goldstone equation. A few years ago, quantum Monte Carlo calculations of single and double $\Lambda$-hypernuclei were also done using two- and three-body forces between the $\Lambda$ and the nucleons \cite{Lonardoni13,Lonardoni14}. The quality of the description of hypernuclei in most of these phenomenological and microscopical approaches relies on the validity of the mean field picture which, however, can substantially change due to the correlations induced by the YN interaction. While many authors have extensively studied the correlations of nucleons in nuclear matter and finite nuclei, those of hyperons have not received so much attention so far. The effect of the $\Lambda$ correlations in nuclear matter, beyond the mean field description, was studied for the first time by Robertson and Dickhoff \cite{Robertson04} using the Green’s function formalism. These authors calculated the spectral function and quasi-particle parameters of the $\Lambda$ finding results qualitatively similar to those of the nucleons. They showed that the 
$\Lambda$ is, in general, less correlated than the nucleons. A few years ago, the spectral function of the $\Lambda$ hyperon in finite nuclei was studied \cite{Vidana17}, showing, in agreement with the work of Robertson and Dickhoff, that the $\Lambda$ is less correlated than the nucleons, and confirming the idea that it maintains its identity inside the nucleus. The results of this study showed also that in hypernuclear production reactions, the $\Lambda$ hyperon is mostly formed in a quasi-free state.

In this work we aim to study the structure of single $\Lambda$-hypernuclei using the Hartree--Fock--Bogoliubov (HFB) method \cite{RingSchuck}. To such end we use finite range Gogny-type interactions to describe both the nucleon-nucleon (NN) and $\Lambda$-nucleon ($\Lambda$N) interactions. In particular, we consider the D1S force \cite{Berger91} for the pure nucleonic sector whereas the unknown parameters of the $\Lambda$-nucleon force are obtained by fitting the experimental binding energies of the $1s$ 
$\Lambda$ single-particle state in various hypernuclei. Then, using the best set of parameters of the $\Lambda$-nucleon interaction, we calculate the binding energies of the other $\Lambda$ single-particle states in the different hypernuclei and compared the predicted results with experimental data. Finally, we also compute the root-mean-square radii of ground state $\Lambda$ orbitals as well as global properties of the hypernuclei such as the pairing energy 
and the quadrupole deformation.

The manuscript is organized in the following way. The application of the HFB method to the study of the structure of single $\Lambda$-hypernuclei is briefly described in Sec.\ \ref{sec:HFB}. The Gogny-type $\Lambda$N force constructed in this work and the fitting procedure followed to determine its parameters is presented in Sec.\ \ref{sec:force}.  Results are shown and discussed in Sec.\ \ref{sec:res}. Finally, a short summary and the main conclusions of this work are given in Sec.\ \ref{sec:suc}.


\section{HFB method for hypernuclei}
\label{sec:HFB}
A general two-body Hamiltonian describing the structure of a single $\Lambda$-hypernucleus reads in second quantization
\begin{equation}
\begin{split}
 H&=\sum_{ij}t^N_{ij}a^\dagger_i a_j + \sum_{ij}t^\Lambda_{ij}c^\dagger_i c_j 
 + \frac{1}{4}\sum_{ijkl}\bar{V}^{NN}_{ij,kl}a^\dagger_ia^\dagger_j a_l a_k \\ &+
    \sum_{ijkl}V^{\Lambda N}_{ijkl}a^\dagger_ic^\dagger_j c_la_k-T_{cm}  \ ,
\end{split}
\label{eq:Hamiltonian2}
\end{equation}
where $a^\dagger_i$ and $c^\dagger_i$ ($a_i$ and $c_i$) are, respectively, the nucleon and $\Lambda$
creation (annihilation) operators, $t^{N(\Lambda)}_{ij}=\bra{i}-\frac{\hbar^2\nabla^2}{2M_{N(\Lambda)}}\ket{j}$  are kinetic energy matrix elements, $\bar{V}^{NN}_{ij,kl}$=$\bra{ij}V_{NN}\ket{kl}-\bra{ij}V_{NN}\ket{lk}$ is the two body matrix element describing the anti-symmetrized NN interaction, and $V^{\Lambda N}_{ijkl}$=$\bra{ij}V_{\Lambda N}\ket{kl}$ is the two body matrix element describing the $\Lambda$N interaction. The last term  is the center-of-mass energy correction taken into account because of the violation of translation invariance in mean field calculations \cite{bender00}.

Starting from this Hamiltonian, the HFB equations can be obtained by minimizing the total ground-state energy of the hypernucleus 
$^{A+1}_{\,\,\,\,\,\,\Lambda}Z$
\begin{equation} \label{eq:Energy}
E=\frac{\bra{\Psi_{^{A+1}_{\,\,\,\,\,\,\Lambda}Z}}H\ket{\Psi_{^{A+1}_{\,\,\,\,\,\,\Lambda}Z}}}
{\bra{\Psi_{^{A+1}_{\,\,\,\,\,\,\Lambda}Z}}\Psi_{^{A+1}_{\,\,\,\,\,\,\Lambda}Z}\rangle}
\end{equation}
where the wave function $\ket{\Psi_{^{A+1}_{\,\,\,\,\,\,\Lambda}Z}}$, 
describing the ground-state of the hypernucleus, is taken as the direct 
product  $\ket{\Psi_{^AZ}}\otimes\ket{\Psi_\Lambda}$ of the nuclear 
core $^AZ$ and the single-particle $\Lambda$ state.
The nuclear part is taken as a Hartree--Fock--Bogoliubov (HFB) wave function.
It is a product wave function defined as the vacuum of quasiparticle annihilation operators
$\beta_{\mu}$, {\it i.e.} $\beta_{\mu} \ket{\Psi_{^AZ}}=0$, where creation and annihilation quasiparticle
operators are defined through the standard Bogoliubov transformation  \cite{RingSchuck,BlaizotRipka}
\begin{equation}\label{eq:Wtransf}
\left(\begin{array}{c}
\beta\\
\beta^\dagger
\end{array}\right)=\left(\begin{array}{cc}
U^{+} & V^{+}\\
V^{T} & U^{T}
\end{array}\right)\left(\begin{array}{c}
a\\
a^\dagger
\end{array}\right)\equiv W^{+ }\left(\begin{array}{c}
a\\
a^\dagger
\end{array}\right) \,,
\end{equation}
expressed in terms of a set of single particle creation and annihilation
operators (the basis) and the $U$ and $V$ amplitudes. 
The label $\mu$ above indexes the quasiparticle 
configurations and contains quantum numbers of preserved symmetries like parity or 
projection of angular momentum along the intrinsic $z$ axis (the $K$ 
quantum number). In the present calculation the basis states are taken
as axially symmetric harmonic oscillator (HO) states, characterized by two
oscillator lengths $b_{\perp}$ and $b_{z}$. The number of states included
in the basis is nucleus dependent and is taken large enough as to guarantee 
the converge of physical quantities  like energies and radii. The Bogoliubov
amplitudes are chosen as to preserve axial symmetry but not reflection symmetry
as to consider octupole effects. 

In the present calculation we are
considering a single hyperon whose wave function is expressed as a 
unitary transformation of a ``hyperon'' basis of the HO type. Both the 
amplitudes of the Bogoliubov transformation and the unitary matrix defining
the hyperon wave function are determined as to minimize the energy  of Eq .\ (\ref{eq:Energy})
according to the variational principle. In the numerical implementation
of the minimization process we use a gradient method with approximate
curvature corrections \cite{second-order-grad}.
The Bogoliubov transformation breaks particle number symmetry and, therefore,
the minimization procedure has to be carried out with a constraint in 
the average number of protons and neutrons in the nucleus wave function. 

Most of the core nuclei considered in the results section are odd mass or 
odd-odd mass nuclei. For a proper and consistent treatment of those systems
in the HFB framework one has to consider ``blocked'' configurations \cite{RingSchuck,BlaizotRipka}
breaking explicitly time-reversal invariance. As a consequence, time-odd
field have to be considered in the calculation of the mean field potential
and pairing field. This represents an additional complication in the proposed
procedure and therefore we have decided to neglect the effects of ``blocking'' and
consider odd and odd-odd mass nuclei as even-even systems with the average number
of protons and neutrons adjusted to the real value of the target system. This is not
a bad approximation as typically the impact of time-odd fields in binding energies
(the most important observable considered here) is minor and can be neglected.

\section{Gogny-type $\Lambda$N force and fitting procedure}
\label{sec:force}

In general, any Gogny nucleon-nucleon interaction can be cast into the form
\begin{eqnarray}
V_{NN}(\vec r)&=&\sum_{i=1}^{2}e^{-\frac{|\vec r|^2}{\mu_i^2}}\left(W_i+B_iP_\sigma-H_iP_\tau-M_iP_\sigma P_\tau \right) \nonumber \\
&+&t_0(1+x_0P_\sigma)\rho^\alpha_N(\vec R)\delta(\vec r) \nonumber \\
&+&i\,W_{LS}\overleftarrow{\nabla}_{12}\delta(\vec r)\times\overrightarrow{\nabla}_{12}\cdot(\vec \sigma_1+\vec \sigma_2)\ , 
\label{eq:GognyNN}
\end{eqnarray}
where $\vec r =\vec r_1 -\vec r_2$, $\vec R=(\vec r_1+\vec r_2)/2$, 
$\nabla_{12}=\nabla_1-\nabla_2$, and $P_\sigma=(1+\vec 
\sigma_1\cdot \vec \sigma_2)/2$ and $P_\tau=(1+\vec \tau_1 \cdot \vec 
\tau_2)/2$ are the spin and isospin exchange operators which give the 
spin-isospin structure of the force. The sum of the two Gaussian-shape 
terms mimic the finite-range effects of a realistic interaction in the 
medium, whereas the last two ones correspond, respectively, to a 
zero-range density-dependent term, that would mimic the effect of 
multi-nucleon forces ({\it e.g.,} NNN, NNNN, ...), and to the 
spin-orbit interaction. In the present work, we use the D1S 
parametrization \cite{Berger91} of the Gogny force which is well known 
for reproducing many  nuclear properties in nuclei in the neighbourhood of  the 
stability region of the Segr\`e Chart of Nuclides. The Coulomb 
repulsion among protons has also been taken into account in the 
traditional way: Coulomb direct field is computed exactly, Coulomb
exchange potential is approximated by the Slater approximation and
Coulomb antipairing is fully neglected \cite{Berger91}.

For the $\Lambda$N interaction we use a very simplified version of the Gogny force, comprising only the central term 
\begin{equation}
\label{hyp-nuc-Gogny}
    V_{\Lambda N}(\vec r) = \sum_{i=1}^{2}e^{-\frac{|\vec r|^2}{\mu_i^2}} (W_i + B_iP_{\sigma}) \ .
\end{equation}
By ignoring the isospin structure of the $\Lambda$N force we assume 
that the $\Lambda$-neutron and $\Lambda$-proton interactions are the 
same. This is a reasonable assumption because the isospin of the $\Lambda$ is 
zero and, therefore, the $\Lambda$N interaction occurs only in the 
total isospin $\frac{1}{2}$ channel. The density-dependent term, 
accounting for the effect of multi-body forces among the $\Lambda$ and 
the nucleons, is also neglected. Finally, the $\Lambda$N spin-orbit 
term is ignored since, as it is well known, the spin-orbit splitting of 
the single-particle levels in $\Lambda$ hypernuclei is very small, 
typically more than one order of magnitude smaller than their nucleonic 
counterparts \cite{Gibson95}. 

The six unknown parameters of the $\Lambda$N interaction, $p=(\mu_1, 
\mu_2, W_1, W_2, B_1,B_2$), are determined by fitting the experimental 
binding energies of the $1s$ $\Lambda$ single-particle state in various 
hypernuclei ($^{16}_{\,\,\Lambda}$O, $^{28}_{\,\,\Lambda}$Si, 
$^{32}_{\,\,\Lambda}$S, $^{40}_{\,\,\Lambda}$Ca, 
$^{51}_{\,\,\Lambda}$V, $^{89}_{\,\,\Lambda}$Y, 
$^{139}_{\,\,\,\,\Lambda}$La, $^{208}_{\,\,\,\,\Lambda}$Pb)  by a 
$\chi^2$ minimization procedure based on the ``Simulated Annealing 
Method'' (SAM) \cite{Kirkpatrik83,Press92}. The SAM technique is based 
on the Metropolis algorithm \cite{Metropolis53} and it is used here to find a global 
minimum that is hidden among many local ones in the hyper-surface of a 
multi-variable function. This method was used for the first time in 
Ref.\ \cite{Agrawal06} to determine the parameters of the Skyrme NN 
interaction, and later in Ref.\ \cite{Guleria12} to find those of a 
Skyrme $\Lambda$N one that was employed  to study the properties of 
single-$\Lambda$ hypernuclei within a Skyrme--Hartree--Fock model. The 
$\chi^2$ function to minimize is defined as 
\begin{equation}
\chi^2=\frac{1}{N_d-N_p}\sum_{i=1}^{N_d}\left(\frac{M^{exp}_i-M^{th}_i}{\sigma_i}\right)^2 \ ,
\label{eq:chi2}
\end{equation}
where $N_d$ and $N_p$ are, respectively, the number of experimental 
data points and the number of parameters used in the fitting, 
$M^{exp}_i$ and $M^{th}_i$ are the experimental and theoretical values 
of a given observable, in our case the binding energies of the 
$\Lambda$ single-particle states, and $\sigma_i$ is its theoretical 
error, taken here equal to the experimental one. 

\begin{table}[t!]
\begin{center}
\begin{tabular}{c|cccc}
\hline
\hline
Parameter & $\Lambda$N$_1$ & $\Lambda$N$_2$ & $\Lambda$N$_3$ & D1S \\
\hline
$\mu_1$ (fm)  & $0.5$ & $0.5$ & $0.7$  & $0.7$ \\ 
$\mu_2$ (fm)  & $1.3$ & $1.4$ & $1.3$ &  $1.2$ \\
$W_1$ (MeV)  & $-157.818$ & $-147.322$ & $-141.499$ & $-1720.30$  \\
$W_2$ (MeV)  & $8.282$    & $8.541$ & $13.579$  & $103.639$\\
$B_1$ (MeV)  & $52.210$  & $42.097$   & $69.193$  & $1300$ \\
$B_2$ (MeV)  & $-32.219$      & $-30.201$ & $-22.993$ & $-163.483$ \\
$\chi^2$          & $35.115$& $42.212$ & $78.938$ &  \\
\hline
\hline
\end{tabular}
\end{center}
\caption{Optimal values of the parameters of the $\Lambda$N Gogny-type 
force obtained from the minimization of the $\chi^2$ function of Eq.\ 
(\ref{eq:chi2}). Three forces have been obtained using three different 
sets of the initial values of the parameters in the SAM. The value of 
$\chi^2$ is shown in the last row. The values of the corresponding 
parameters for the D1S NN interaction are also shown for comparison in 
the last column.}
\label{tab:tab1}
\end{table}

The SAM consist on three steps. In the first one, the binding energies 
of the $1s$ $\Lambda$ single-particle state of various hypernuclei are 
calculated for an initial guess of the unknown parameters $p$ which 
together with the experimental binding energies allow to determine 
$\chi^2$. Then, in the second step the value of one of the parameters is 
randomly chosen and modified according to
\begin{equation}
p \rightarrow p +\eta \Delta p \ ,
\label{eq:parametermod}
\end{equation}
where $\Delta p$ is the maximum displacement allowed in a single step 
for a given parameter $p$, and $\eta$ is a random number uniformly 
distributed between $-1$ and $1$. This second step is repeated until 
the obtained new values of the parameters are found to be within a 
range of values $(p_{min},p_{max})$ which is defined at the beginning 
of the procedure. Once this condition is satisfied, $\chi^2$ is 
then calculated again for the new set of parameters.  

Finally, in the last step of the SAM, the new generated set of 
$\Lambda$N interaction parameters is accepted or rejected with the help 
of the Metropolis algorithm. To such end we calculate the quantity
\begin{equation} 
P(\chi^2)=exp\left(\frac{\chi^2_{old}-\chi^2_{new}}{T}\right) \ ,
\label{eq:prob}
\end{equation}
where $\chi^2_{old}$ and $\chi^2_{new}$ are, respectively, the values 
of $\chi^2$ obtained with the initial guess of parameters and the final 
one obtained at the end of the second step. $T$ is a control parameter 
which, in analogy with the way liquids freeze or metals 
recrystallize in the annealing process in which the SAM is inspired, 
is usually referred to as effective temperature. 

The new set of parameters is accepted only if $P(\chi^2)>\beta$ where $\beta$ is a random number uniformly distributed between $0$ and $1$. Accepted sets of parameters are called ``successful''. 

Starting from an initial guess $T_{in}$  of the control parameter steps 
2 and 3 are repeated until we obtain $100N_p$ new sets of parameters or 
$10N_p$ successful sets, whichever happens first. Once this is 
achieved, the value of the control parameter $T$ is then reduced by 
using a suitable annealing schedules and the entire process of 
minimization of $\chi^2$ is carried out again. The reduction of $T$ is 
done until further minimization of $\chi^2$ becomes unnecessary. Of the 
several annealing schedules available in the literature
we have chosen the Cauchy one \cite{Cohen94} in this work. It is  given by
\begin{equation}
T_i=\frac{T_{in}}{ck} \ , \,\,\, k=1,2,3,...
\label{eq:cauchy}
\end{equation}
where $c$ is a constant taken equal to $1$ in this work . 

The optimal values of the parameters of the $\Lambda$N Gogny-type force 
obtained from the minimization of the $\chi^2$ function are given in 
Tab.\ \ref{tab:tab1}. Since the SAM starts with an initial guess of the 
six parameters, we present the parameters for three forces obtained 
using three different initial guesses of their values. For comparison 
the values of the corresponding parameters for the D1S NN interaction 
are also shown in the last column. Note that whereas the range of the 
Gaussian functions ($\mu_i$) found for the three forces is similar to 
that of the D1S NN force, the values of parameters $W_i$, $B_i$ that 
characterize the strength of the $\Lambda$N force are much more smaller 
in absolute value than their nucleonic counterparts. This is in 
agreement with the idea that the $\Lambda$N interaction is weaker than 
the NN one. 

\section{Results and Discussion}
\label{sec:res}

\begin{table}[t!]
\begin{center}
\begin{tabular}{c|ccccc}
\hline
\hline
 &  & $\Lambda$N$_1$ & $\Lambda$N$_2$ & $\Lambda$N$_3$ & Exp. \\
\hline
$^{16}_{\,\,\Lambda}$O  & $1s$ & $11.90$ & $11.41$ & $13.29$ & $13.0(2)$ \\ 
                                       & $1p$ & $0.99$ & $0.84$ & $1.34$ & $2.5(2)$ \\
\\
$^{28}_{\,\,\Lambda}$Si  & $1s$ & $17.81$ & $17.35$ & $18.68$ & $17.2(2)$ \\
                                        & $1p$ & $6.21$ & $5.94$ & $6.65$ & $7.6(2)$ \\
\\
$^{32}_{\,\,\Lambda}$S  & $1s$ & $20.11$ & $19.62$ & $21.05$ & $17.5(5)$ \\
                                       & $1p$ & $8.27$ & $7.96$ & $8.79$ & $8.2(5)$ \\
 \\ 
$^{40}_{\,\,\Lambda}$Ca  & $1s$ & $19.66$ & $19.32$ & $19.98$ & $18.7(1.1)^\dag$ \\
                                        & $1p$ & $9.57$ & $9.32$ & $9.82$ & $11.0(5)$ \\
                                         & $1d$ & $0.37$ & $0.26$ & $0.37$ & $1.0(5)$ \\
\\                                         
$^{51}_{\,\,\Lambda}$V  & $1s$  & $21.43$ & $21.17$ & $21.39$ & $21.5(6)$ \\
                                       & $1p$ & $11.93$ & $11.70$ & $12.02$ & $13.4(6)$ \\
                                       & $1d$ & $2.55$ & $2.42$ & $2.52$ & $5.1(6)$ \\
\\
$^{89}_{\,\,\Lambda}$Y  & $1s$  & $24.10$ & $23.98$ & $23.48$ & $23.6(5)$ \\
                                       & $1p$ & $16.98$ & $16.83$ & $16.64$ & $17.7(6)$ \\ 
                                       & $1d$ & $9.40$ & $9.25$ & $9.23$ & $10.9(6)$ \\
                                       & $1f$ & $1.69$ & $1.58$ & $1.55$ & $3.7(6)$ \\
\\
$^{139}_{\,\,\,\,\Lambda}$La  & $1s$  & $25.22$ & $25.19$ & $24.30$ & $25.1(12)$ \\
                                               & $1p$ & $19.79$ & $19.71$ & $19.16$ & $21.0(6)$ \\ 
                                               &$1d$ & $13.71$ & $13.60$ & $13.30$ & $14.9(6)$ \\
                                               & $1f$ & $7.16$ & $7.04$ & $6.88$ & $8.6(6)$ \\
                                               & $1g$ & $0.36$ & $0.26$ & $0.12$ & $2.1(6)$ \\
 \\                                              
$^{208}_{\,\,\,\,\Lambda}$Pb  & $1s$  & $26.20$ & $26.22$ & $25.12$ & $26.9(8)$ \\
                                               & $1p$ & $21.86$ & $21.83$ & $21.01$ & $22.5(6)$ \\ 
                                               &$1d$ & $16.94$ & $16.87$ & $16.29$ & $17.4(7)$ \\
                                               & $1f$ & $11.49$ & $11.20$ & $11.00$ & $12.3(6)$ \\
                                               & $1g$ & $5.63$ & $5.53$ & $5.25$ & $7.2(6)$ \\
\hline
\hline
\end{tabular}
\end{center}
\caption{Binding energy of $\Lambda$ single-particle states in the eight hypernuclei considered for the three $\Lambda$N Gogny-type forces constructed. Experimental values are taken from the compilation of 
Ref.\ \cite{Gal16}. $^\dag$The weak signal of $^{40}_{\,\,\Lambda}$Ca \cite{Pile91} is not included in this compilation. Units are given in MeV.}
\label{tab:tab2}
\end{table}

\begin{figure}[t!]
\begin{center}
\vskip -0.cm
\includegraphics[width=6.8cm,angle=0,clip]{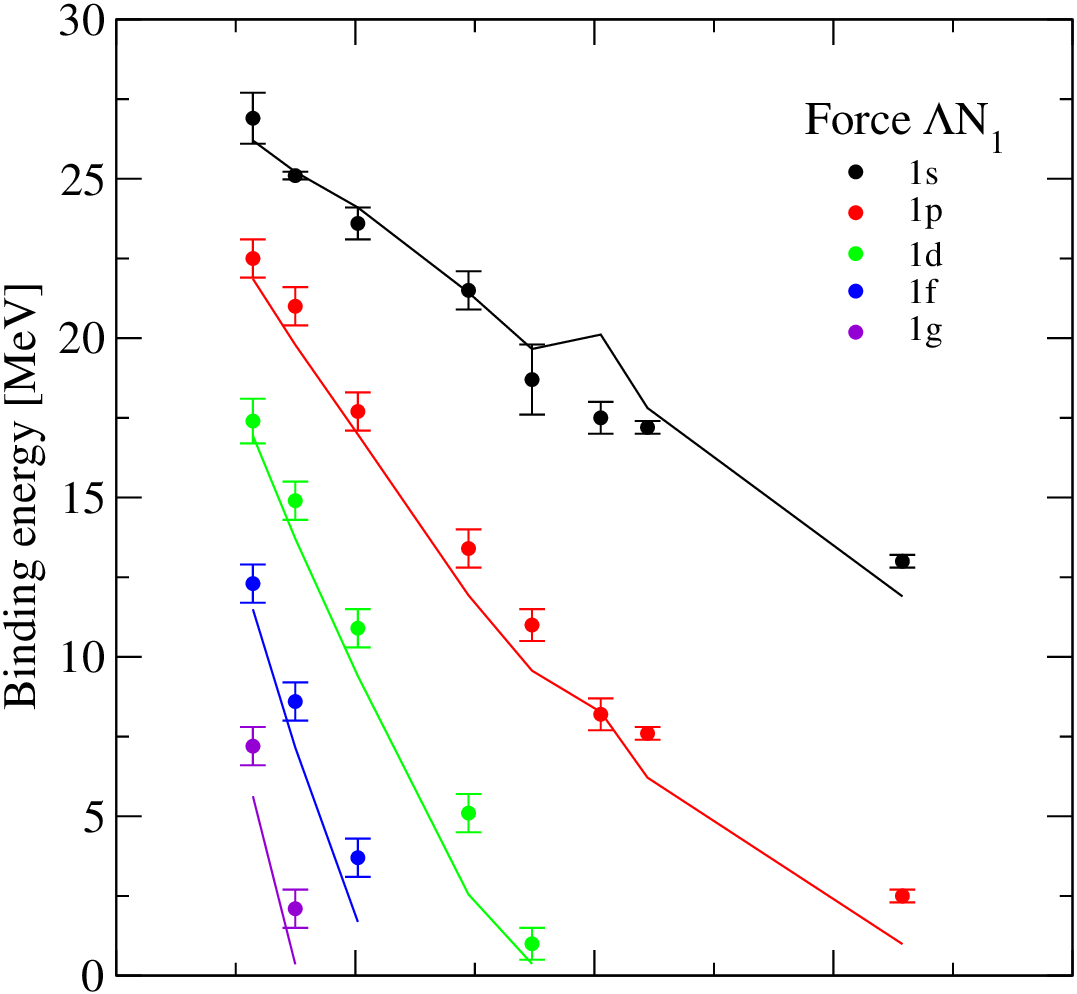}
\vskip -0.cm
\includegraphics[width=6.8cm,angle=0,clip]{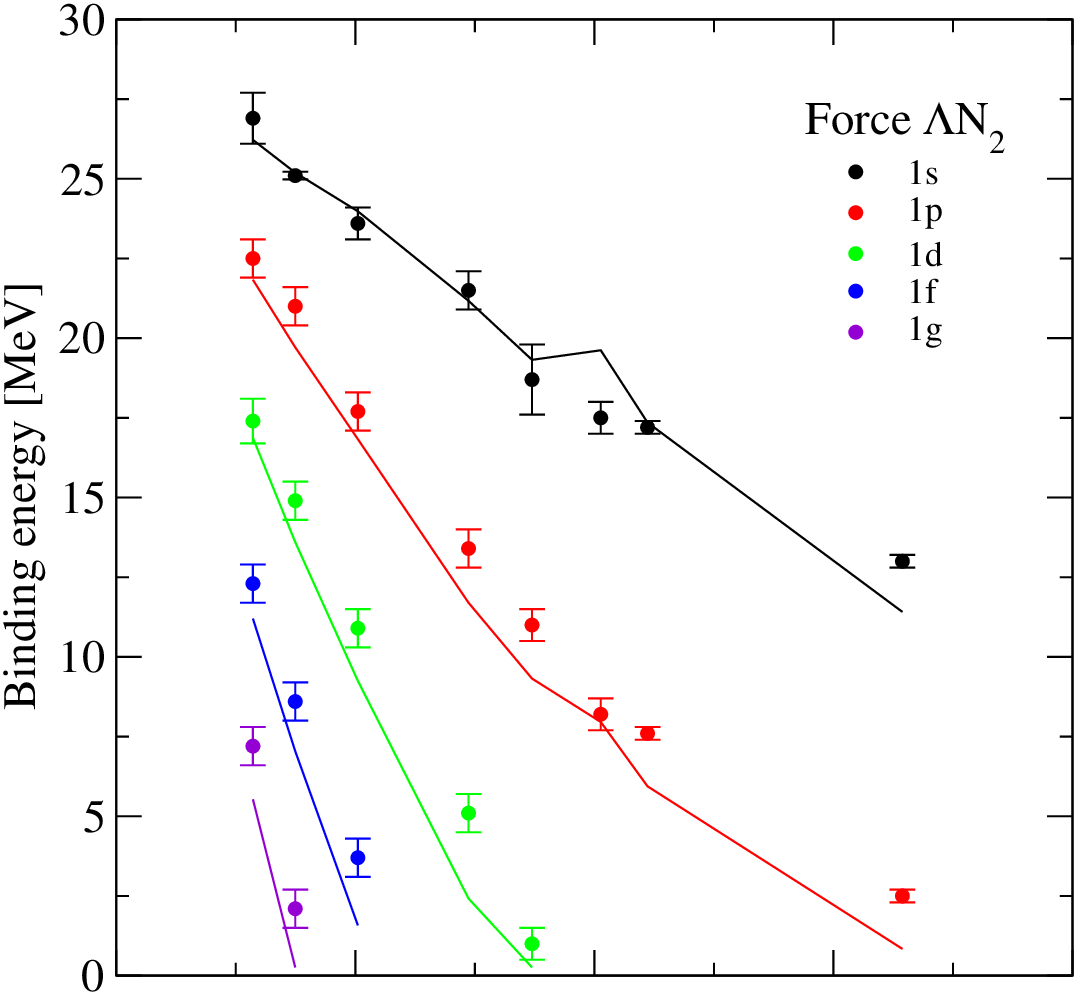}
\vskip -0.cm
\hskip 0.2cm
\includegraphics[width=7.0cm,angle=0,clip]{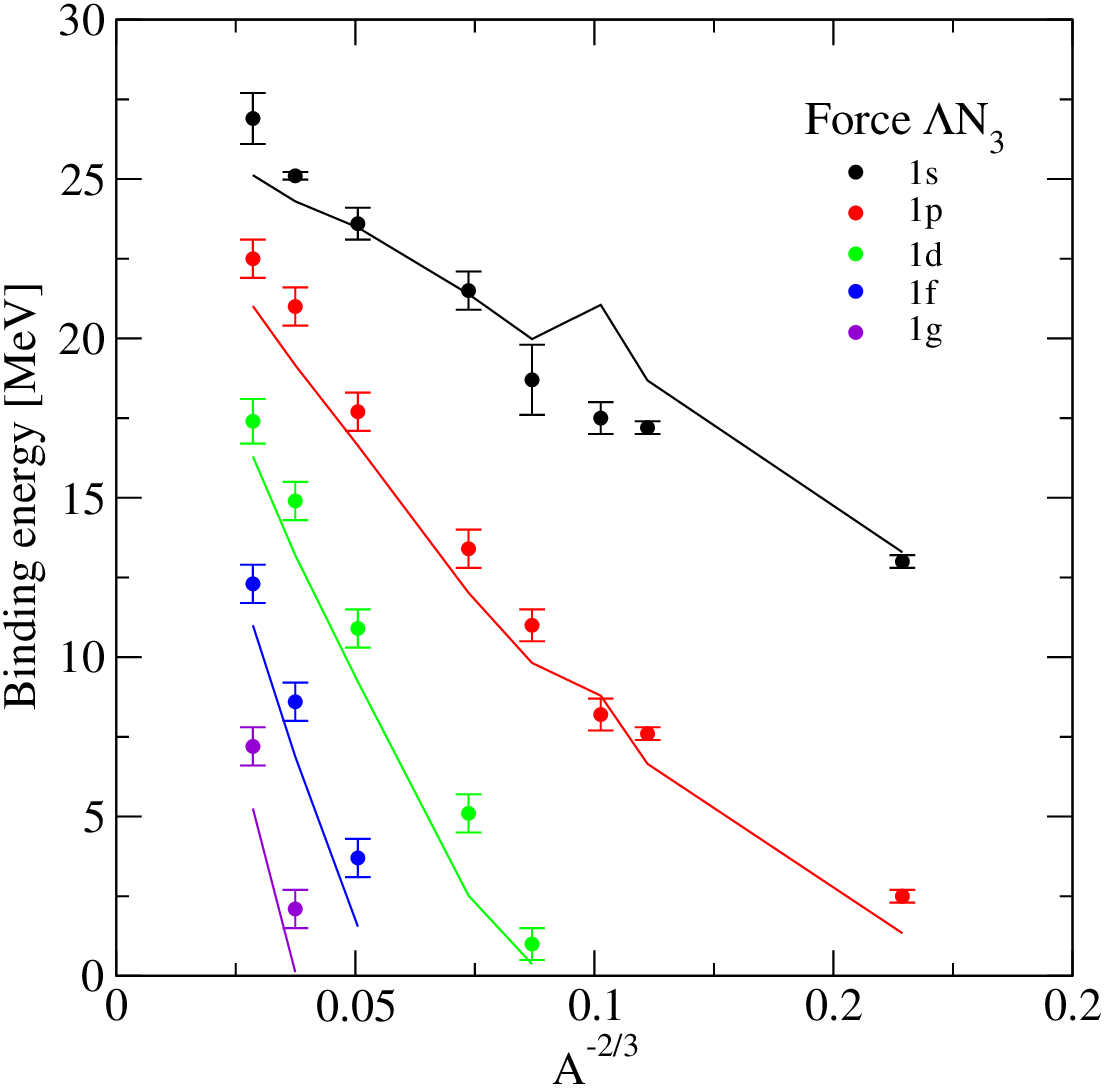}
\caption{Binding energies of the $1s$, $1p$, $1d$, $1f$, and $1g$ $\Lambda$
single-particle states as a function of A$^{-2/3}$. Experimental values are
taken from the compilation of Ref. \cite{Gal16}. 
}
\label{fig:fig1}
\end{center}
\vskip -0.5cm
\end{figure}

The binding energy of the $\Lambda$ single-particle states in 
$^{16}_{\,\,\Lambda}$O, $^{28}_{\,\,\Lambda}$Si, 
$^{32}_{\,\,\Lambda}$S, $^{40}_{\,\,\Lambda}$Ca, 
$^{51}_{\,\,\Lambda}$V, $^{89}_{\,\,\Lambda}$Y, 
$^{139}_{\,\,\,\,\Lambda}$La and $^{208}_{\,\,\,\,\Lambda}$Pb are shown 
in Tab.\ \ref{tab:tab2} for the three $\Lambda$N Gogny-type forces 
constructed together with the experimental values taken from the 
compilation made by Gal {\it et al.,} in Ref.\ \cite{Gal16}. A graphic 
representation of our results is provided in Fig.\ \ref{fig:fig1}, 
where the binding energies of the different $\Lambda$ single-particle 
states are shown as a function of $A^{-2/3}$ with $A=N+Z$ the mass 
number of the nuclear core $^AZ$ of the hypernucleus 
$^{A+1}_{\,\,\,\,\,\,\Lambda}Z$. As it can be seen in the table, and 
more clearly in the figure, the forces $\Lambda$N$_1$ and 
$\Lambda$N$_2$ fit with better accuracy the binding energy of the $1s$ 
$\Lambda$ single-particle state than the $\Lambda$N$_3$ one. We note, 
however, that the three of them clearly overbind the $1s$ state of 
$^{32}_{\,\,\Lambda}$S. A possible explanation for the overbinding of 
this state is that the ground-state shape of $^{31}$S (that forms the nuclear 
core of $^{32}_{\,\,\Lambda}$S) is not well defined in our calculation. 
The potential energy surface predicted by the Gogny D1S NN interaction 
shows a very flat ground state minimum as a function of the 
quadrupole deformation parameter $\beta_2$ (see Refs.\ 
\cite{Hilaire07,CEA}), that extends from the oblate  
to the prolate  region. The overbinding of the $1s$ state 
of $^{32}_{\,\,\Lambda}$S could be probably reduced if the shape of 
$^{31}$S were determined in our calculation with less uncertainty
by introducing beyond mean field fluctuations in the quadrupole degree of freedom \cite{review}.
This improvement is, however, out of the scope of the present work. 
Regarding the $1p, 1d, 1f$ and $1g$ single-particle states, which have 
not been explicitly included in the fitting procedure, the three 
$\Lambda$N forces predict their binding energies in reasonable good 
agreement with the experimental data, although a slight underbinding of 
these states is observed in all hypernuclei. 

\begin{figure}[t!]
\begin{center}
\includegraphics[width=6.8cm,angle=0,clip]{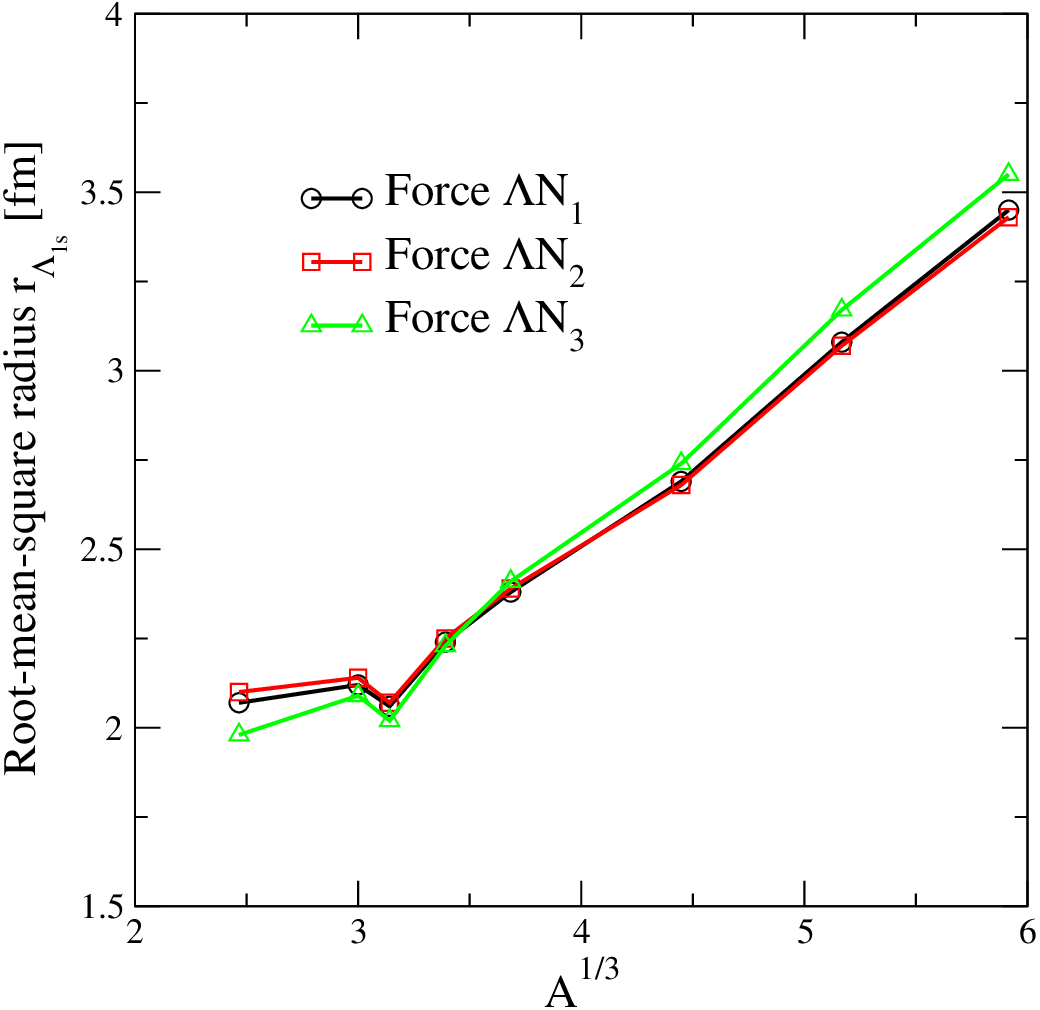}
\caption{Root-mean-square radius of the $1s$ $\Lambda$
single-particle state as a function of A$^{1/3}$.}
\label{fig:fig2}
\end{center}
\end{figure}

\begin{table*}[t!]
\begin{center}
\scriptsize
\begin{tabular}{c|ccc|ccc|ccc|c|cccc}
\hline
\hline
& \multicolumn{3}{c}{$\Lambda$N$_1$} & \multicolumn{3}{c}{$\Lambda$N$_2$} & \multicolumn{3}{c}{$\Lambda$N$_3$} & \multicolumn{4}{c}{} \cr
\hline 
& $E_{HFB}$ & $\Delta$  & $Q$ & $E_{HFB}$ & $\Delta$  & $Q$  & $E_{HFB}$ & $\Delta$   & $Q$  && $E_{HFB}$ & $\Delta$   & $Q$     \\
\hline
$^{16}_{\,\,\Lambda}$O  & $-145.16$ & $-1.56$ & $-1.9\times 10^{-5}$& $-144.09$  & $-1.57$ & $-2.1\times 10^{-5}$& $-148.01$ & $-1.51$ & $-1.6\times 10^{-5}$ & $^{15}$O & $-114.96$ & $-1.75$ & $-1.4\times 10^{-4}$  \\
$^{28}_{\,\,\Lambda}$Si & $-258.56$ & $-3.30$ & $18.70$& $-257.68$  & $-3.27$ &$18.52$ & $-260.06$ & $-3.41$ & $19.74$ & $^{27}$Si & $-216.93$ & $-3.27$  & $25.34$ \\
$^{32}_{\,\,\Lambda}$S  & $-301.11$ & $-1.19$ & $2.3\times 10^{-3}$&  $-300.14$ & $-1.20$ & $1.7\times 10^{-3}$& $-302.76$ & $-1.18$ & $5\times 10^{-3}$ 
& $^{31}$S & $-254.39$ & $-2.83$ & $0.15$ \\
$^{40}_{\,\,\Lambda}$Ca & $-374.33$ & $-2.61$ & $4\times 10^{-6}$& $-373.69$ & $-2.61$ & $4\times 10^{-6}$ & $-374.75$ & $-2.58$ & $5\times 10^{-6}$ 
& $^{39}$Ca & $-329.34$ & $-2.87$ & $-1 \times 10^{-6}$ \\
$^{51}_{\,\,\Lambda}$V   & $-484.35$ & $-6.27 $ &$49.62$ & $-483.88$ & $-6.30 $ & $49.01$ & $-484.01$ & $-6.17$ & $51.82$ 
& $^{50}$V & $-436.40$ & $-8.76$ & $-3.00$ \\
$^{89}_{\,\,\Lambda}$Y   & $-819.31$ & $-8.53$ & $-0.15 $& $-819.13$ & $-8.54$ & $-0.14$ & $-817.88$ & $-8.52$ & $-0.14$ 
& $^{88}$Y & $-767.17$ & $-8.88$ & $-0.15$ \\
$^{139}_{\,\,\,\,\Lambda}$La  & $-1210.10$ & $-17.08 $ & $-0.32$& $-1210.08$ & $-17.08$ & $-0.40$& $-1208.13$ & $-17.09$ & $-0.25$ 
& $^{138}$La & $-1156.74$ & $-16.76$ & $-0.56$  \\
$^{208}_{\,\,\,\,\Lambda}$Pb &$-1684.69$ & $-2.63$ & $-0.06$& $-1684.76$ & $-2.63 $ & $-0.06$ & $-1682.42$ & $-2.62$ & $-0.06$ 
& $^{207}$Pb & $-1629.76$ & $-2.27$ & $-0.40$\\
\hline
\hline
\end{tabular}
\end{center}
\caption{Ground-state HFB energy, pairing energy $\Delta$ and quadrupole moment $Q$ of the hypernuclei considered. The results
of the corresponding ordinary nuclei forming the nuclear core of each hypernuclei are shown for comparison in the last three columns. HFB and pairing energies are given in MeV whereas the quadrupole moment is given in fm$^2$.}
\label{tab:tab3}
\end{table*}

We show now in Fig.\ \ref{fig:fig2} the root-mean-square radius of the $1s$ $\Lambda$ single-particle state as a function of $A^{1/3}$ predicted by the three $\Lambda$N forces, defined as
\begin{equation}
r_{\Lambda_{1s}}=\sqrt{\frac{\langle\Psi_{\Lambda_{1s}}|\hat r^2|\Psi_{\Lambda_{1s}}\rangle}
{\langle \Psi_{\Lambda_{1s}}|\Psi_{\Lambda_{1s}}\rangle}} \ ,
\label{eq:rmsr}
\end{equation}
with $\Psi_{\Lambda_{1s}}$ the wave function of the $1s$  $\Lambda$ single-particle state.

We note first that the three models give similar values of 
$r_{\Lambda_{1s}}$ and that they are in good agreement with those 
reported in other works such as for instance those of Refs.\ 
\cite{skhf4,skhf5,skhf5c,skhf8,Guleria12} where single 
$\Lambda$-hypernuclei were studied within a Skyrme--Hartree--Fock model 
or the one of Ref.\  \cite{Vidana17} where hypernuclear structure was 
treated within a perturbative many-body approach using meson exchange 
based hyperon-nucleon interactions. We note also that the lower the 
binding energy (see Tab.\ \ref{tab:tab2} ), the larger the 
root-mean-square radius. The exception is $^{32}_{\,\,\Lambda}$S for 
which the three $\Lambda$N forces predict a value of $r_{\Lambda_{1s}}$ 
smaller than that of the $1s$ $\Lambda$ single-particle state in 
$^{28}_{\,\,\Lambda}$Si. The reason is the overbinding of this state 
which, as mentioned before, could probably be ascribed to the uncertain 
shape of the nuclear core $^{31}$S predicted by the D1S NN interaction 
employed in this work. The increase of $r_{\Lambda_{1s}}$ with the 
nuclear mass observed is a direct consequence of the spreading of the 
$1s$ state due to the larger spatial extension of the nuclear density 
over which the $\Lambda$ hyperon wants to be distributed when going 
from the lighter to the heavier hypernuclei. That is, the wave function 
of the $1s$ $\Lambda$ single-particle state is more extended when 
increasing the nuclear core of the hypernucleus and, consequently, 
$r_{\Lambda_{1s}}$ increases, showing an approximately linear 
dependence on $A^{1/3}$ in the region from medium to heavy hypernuclei. 

In Tab.\ \ref{tab:tab3} we present now the results for the ground-state 
HFB energy $E_{HFB}$, pairing energy $\Delta$ and quadrupole moment $Q$ 
of the hypernuclei considered. The $\Lambda$ is assumed to be in the 
$1s$ single-particle state. The value of these quantities for the 
corresponding ordinary nuclei forming the nuclear core of each one of 
the hypernuclei are shown for comparison in the last three columns of 
the table. We notice first that, as expected, the additional attraction 
felt by neutrons and protons due to their interaction with the 
$\Lambda$ leads to a clear decrease of ground-state HFB energy of the 
hypernuclei with respect to that of the corresponding nuclear core. We 
note also that the difference between the ground state HFB energy of 
the hypernuclei and that of the nuclear core increases when the size of 
the hypernuclei increases. This is simply because the attraction 
between the $\Lambda$ and the nucleons increases with the density of 
the nuclear core which increases when going from the lightest to the 
heaviest one. 

Regarding the pairing energy, we should mention first that the 
$\Lambda$-nucleon pairing has been ignored in the present calculation 
and that only the nucleon-nucleon pairing has been taken into account. 
However, we observe that the presence of the $\Lambda$ hyperon still 
modifies the pairing energy of the hypernuclei with respect to the 
value that is obtained for the corresponding ordinary nuclei forming 
their nuclear cores. This modification is in the range of $0-0.4$ MeV 
for the majority of the hypernuclei, being the exceptions 
$^{32}_{\,\,\Lambda}$S and $^{51}_{\,\,\Lambda}$V which show a 
difference in their pairing energies with respect to those of $^{31}$S 
and $^{50}$V of about $1.65$ MeV and $2.5$ MeV, respectively. Although 
the $\Lambda$-nucleon pairing has been ignored, the modification of the 
hypernuclei pairing energy should still be attributed to the effect of 
the $\Lambda$ hyperon which affect the solution of the HFB equations of 
the nucleons.

Finally, we observe that the quadrupole moment of the hypernuclei 
gets reduced with respect to that of the ordinary nuclei, being the former 
ones less deformed. The reason can be likely attributed to the fact 
that since the $\Lambda$ hyperon is in the $1s$ state, the probability 
density of finding it at a given position inside the hypernucleus, 
$|\Psi_{\Lambda_{1s}}(r)|^2$, is spherically symmetric and larger at 
the center of the hypernucleus. Therefore, neutrons and protons are 
attracted isotropically towards the center of the hypernucleus making 
it slightly less deformed. There is an exception to this behaviour and
corresponds to the $^{51}_{\,\,\Lambda}$V case. As in the case of 
$^{31}$S (core of $^{32}_{\,\,\Lambda}$S), the potential 
energy of $^{50}$V (core of $^{51}_{\,\,\Lambda}$V)  
around the spherical ground state is very flat extending from the
oblate to the prolate side. The polarization effect of the $\Lambda$ is
enough as to favor a prolate minimum in the $\Lambda$ hypernucleus instead
of the the oblate one observed in the pure nuclear system. As discussed in 
the $^{31}$S case a proper treatment of the problem would require going
beyond the mean field \cite{review} which, however, is out of the scope of the 
present paper and is left for a future work.

\section{Summary and Conclusions}
\label{sec:suc}

In this work we have studied the structure of single $\Lambda$-hypernuclei using the HFB method. To such end
we have used finite range Gogny-type forces to describe both the nucleon-nucleon and $\Lambda$-nucleon interactions. We have built three different 
$\Lambda$-nucleon Gogny forces. The unknown parameters of these forces have been obtained by fitting the experimental binding energies of the $1s$ $\Lambda$ single-particle state in various hypernuclei using the so-called ``Simulated Annealing Method''. These forces have then be used to predict the binding energies of the other $\Lambda$ single-particle states in the different hypernuclei considered. The predicted values have been found to be in good agreement with the experimental data for the three forces constructed. We have also calculated the root-mean-square radii of ground state $\Lambda$ orbital finding 
results in good agreement with those of previous works using different approaches to study the properties of single $\Lambda$-hypernuclei.
Finally we have analyzed several global properties of the hypernuclei considered such as their ground-state HFB energy, their pairing energy and their quadrupole moment and compared them with the ones of the corresponding ordinary nuclei forming the core of the hypernuclei. The ground-state HFB energy was found to be smaller than those of the corresponding ordinary nuclei, the reason being the additional attraction felt by the nucleons due to their interaction with the $\Lambda$ hyperon. Although the $\Lambda$-nucleon pairing was ignored, the pairing energies of hypernuclei were observed to be also modified due to the effect of the $\Lambda$ on the solution of the HFB equations of the nucleons. Finally,  the quadrupole moment of the hypernuclei was found to be reduced with respect to that of the ordinary nuclei, being therefore hypernuclei less deformed. The reason was attributed to the fact since the $\Lambda$ in the ground-state of the hypernuclei is in the $1s$ state, the probability density of finding it at a given position inside the hypernucleus, $|\Psi_{\Lambda_{1s}}(r)|^2$, is spherically symmetric and larger at the center of the hypernuclei. Therefore, neutrons and protons are attracted isotropically towards the center of the hypernuclei making in this way the hypernucleus slightly less deformed. 

\section*{Acknowledgements}

This work is dedicated to the memory of Prof. Peter Schuck.
C.V.N.K. acknowledges the Erasmus Mundus Master on Nuclear Physics (Grant Agreement No. 2019-2130) supported by the Erasmus+ Programme of the European Union for a scholarship. The work of L.M.R. was supported by Spanish Agencia Estatal de Investigaci\'{o}n (AEI) of the Ministry of Science and Innovation under Grant No. PID2021-127890NB-I00.  I.V. thanks the support of the European Union’s Horizon 2020 research and innovation programme under Grant Agreement No. 824093.




\end{document}